\documentclass[a4paper]{PoS}

\usepackage{amsmath}
\usepackage{cite}
\usepackage{sidecap}

\newcommand{\be}{\begin{equation}}
\newcommand{\ee}{\end{equation}}
\newcommand{\bea}{\begin{eqnarray}}
\newcommand{\eea}{\end{eqnarray}}

\title{HVP contribution of the light quarks to the muon $(g - 2)$ including isospin-breaking corrections with Twisted-Mass fermions}

\ShortTitle{$a_\mu^{HVP} (ud)$ including IB corrections}

\author{\speaker{D.~Giusti}$^{(a,b)}$, V.~Lubicz$^{(a,b)}$, G.~Martinelli$^{(c)}$, F.~Sanfilippo$^{(b)}$, S.~Simula$^{(b)}$, and C.~Tarantino$^{(a,b)}$

\\

\it $^{(a)}$ Dipartimento di Matematica e Fisica, Universit\`a  degli Studi Roma Tre, Rome, Italy.\\ Email: \email{davide.giusti@uniroma3.it}, \email{vittorio.lubicz@uniroma3.it}, \email{tarantino@fis.uniroma3.it}

\it $^{(b)}$ Istituto Nazionale di Fisica Nucleare, Sezione di Roma Tre, Rome, Italy.\\ Email: \email{francesco.sanfilippo@infn.it}, \email{simula@roma3.infn.it}

\it $^{(c)}$ Dipartimento di Fisica, Universit\`a  degli Studi di Roma "La Sapienza" and INFN, Sezione di Roma, Rome, Italy.\\ Email: \email{guido.martinelli@roma1.infn.it}

}

\abstract{We present a preliminary lattice calculation of the leading-order electromagnetic and strong isospin-breaking corrections to the Hadronic Vacuum Polarization (HVP) contribution of the light quarks to the anomalous magnetic moment of the muon.
The results are obtained in the quenched-$QED$ approximation using the $QCD$ gauge configurations generated by the European Twisted Mass Collaboration (ETMC) with $N_f = 2 + 1 + 1$ dynamical quarks, at three values of the lattice spacing varying from $0.089$ to $0.062 ~ \mbox{fm}$, at several lattice volumes and with pion masses in the range $M_\pi \simeq 220 \div 490 ~ \mbox{MeV}$.}

\FullConference{The 36th Annual International Symposium on Lattice Field Theory - LATTICE2018\\
		22-28 July 2018\\
		Michigan State University, East Lansing, Michigan, USA.}

\begin{document}

\section{Introduction}
\label{sec:intro}

The muon anomalous magnetic moment $a_\mu \equiv (g - 2) / 2$ is one of the most precisely determined quantities in particle physics.
It is experimentally known with an accuracy of $0.54$ ppm (BNL E821) and the current precision of the Standard Model (SM) prediction is at the level of $0.4$ ppm~\cite{PDG}.
The discrepancy between the experimental value, $a_\mu^{exp}$, and the SM prediction, $a_\mu^{\rm SM}$, corresponds to $\simeq 3.7$ standard deviations, according to the most recent determination of the HVP contribution \cite{Keshavarzi:2018mgv}, namely $a_\mu^{exp} - a_\mu^{\rm SM} = 27.1 ~ (7.3) \cdot 10^{-10}$.
The forthcoming $g - 2$ experiments at Fermilab (E989) and J-PARC (E34) aim at reducing the experimental uncertainty by a factor of four, down to 0.14 ppm, making the comparison of $a_\mu^{exp}$ with the theoretical predictions one of the most stringent tests of the SM in the quest of new physics effects.
With such a reduced experimental error, the uncertainty of the hadronic corrections, due to the HVP and hadronic light-by-light contributions, will soon become the main limitation of this SM test. 

The theoretical predictions for the HVP contribution, $a_\mu^{\rm HVP}$, have been traditionally obtained from experimental data using dispersion relations for relating the HVP function to the experimental cross section data for $e^+ e^-$ annihilation into hadrons~\cite{Davier:2010nc,Hagiwara:2011af}. 
An alternative approach, originally proposed in Ref.~\cite{Lautrup:1971jf}, is to compute $a_\mu^{\rm HVP}$ in lattice $QCD$ from the Euclidean correlation function of two electromagnetic (em) currents. 
An impressive progress in the lattice determinations of $a_\mu^{\rm HVP}$ has been achieved in the last few years, based on improved evaluation of the leading-order hadronic contribution, which is of order ${\cal O} (\alpha^2_{em})$, as well as of the next-to-leading-order hadronic corrections, which include ${\cal O} (\alpha^3_{em})$ contributions.

With the increasing precision of the lattice calculations, it becomes mandatory to include em and strong isospin-breaking (IB) corrections to the HVP, contributing at order ${\cal O} (\alpha^3_{em})$ and ${\cal O} (\alpha^2_{em} (\widehat{m}_d - \widehat{m}_u))$~\footnote{Throughout this contribution by $\widehat{m}$ we indicate a quark mass renormalized in the full theory with both $QCD$ and $QED$ interactions switched on.} respectively.
In Ref.~\cite{Giusti:2017jof} a lattice calculation of both the leading and the IB corrections to the HVP contribution due to strange- and charm-quark intermediate states was carried out, using the time-momentum representation for $a_\mu^{\rm HVP}$~\cite{Bernecker:2011gh} and the expansion method of the path integral in powers of the electromagnetic coupling $\alpha_{em}$ and of the $d$- and $u$-quark mass difference $(\widehat{m}_d - \widehat{m}_u)$ (the RM123 approach of Refs.~\cite{deDivitiis:2011eh,deDivitiis:2013xla}).

In this contribution we present preliminary results of a lattice calculation of the IB corrections to the HVP contribution due to light $u$- and $d$-quark (connected) intermediate states, using the RM123 approach. Given the observed large statistical fluctuations, we do not have yet results for the disconnected contributions.

\section{Time-momentum representation}
\label{sec:TMR}

Following our previous work~\cite{Giusti:2017jof}, we adopt the time-momentum representation for the evaluation of the IB corrections $\delta a^{\rm HVP}_\mu (ud)$ to the $u$- and $d$-quark contributions to the muon anomalous magnetic moment, namely
\be
\delta a^{\rm HVP}_\mu (ud) = 4 \alpha_{em}^2 \int_0^\infty dt ~ f (t) ~ \delta V^{ud} (t) ~ ,
\label{eq:delta_amu_t}
\ee
where $t$ is the Euclidean time, $f (t)$ is a known kinematical kernel~\cite{Bernecker:2011gh} (depending also on the muon mass $m_\mu$) and $\delta V^{ud} (t)$ (see Section~\ref{sec:IB}) represents the IB corrections to the light-quark vector current-current Euclidean correlator $V^{ud} (t)$, defined as
\be
V^{ud} (t) \equiv \sum_{f = u, d} ~ \frac{1}{3} \sum_{i=1,2,3} ~ \int d\vec{x} ~ \langle J_i^f(\vec{x}, t) J_i^f(0) \rangle ~ .
\label{eq:VV}
\ee
In Eq.~(\ref{eq:VV}) $J_\mu^f (x) \equiv  q_f ~ \overline{\psi}_f (x) \gamma_\mu \psi_f (x)$ is the em current with $q_f$ being the electric charge of the quark with flavor $f$ in units of $e$.

A convenient procedure relies on splitting Eq.~(\ref{eq:delta_amu_t}) into two contributions corresponding to $0 \leq t \leq T_{data}$ and $t > T_{data}$, respectively.
In the first contribution the IB corrections to the light-quark vector correlator are directly given by the lattice data, while for the second contribution an analytic representation is required and justified for lattice simulations at non-physical pion masses (see Refs.~\cite{DellaMorte:2017dyu,Giusti:2017jof,Giusti:2018mdh}).
If $T_{data}$ is large enough that the ground-state contribution is dominant for $t > T_{data}$ and it is smaller than $T / 2$ in order to avoid backward signals, one can write
\be
\delta a_\mu^{\rm HVP} (ud) = 4 \alpha_{em}^2 \Biggl\{ \sum_{t = 0}^{T_{data}} f (t) ~ \delta V^{ud} (t)
                                              + \sum_{t = T_{data} + a}^\infty f (t) ~ \frac{Z^{ud}_V} {2 M^{ud}_V} ~ e^{- M^{ud}_V t}
                                              \left[ \frac{\delta Z^{ud}_V}{Z^{ud}_V} - \frac{\delta M^{ud}_V}{M^{ud}_V} (1 + M^{ud}_V t) \right] \Biggr\} ~ .
\label{eq:decomposition}
\ee
In Eq.~(\ref{eq:decomposition}) $M_V^{ud}$ is the ground-state vector-meson mass and $Z_V^{ud}$ is the squared matrix element of the vector current between the state $| V \rangle$ and the vacuum, i.e. $Z_V^{ud} \equiv (1/3) \sum_{i=1,2,3} \sum_{f=u,d} q_f^2 | \langle 0 | \overline{\psi}_f(0) \gamma_i ~ \cdot$ $\cdot ~ \psi_f(0) | V \rangle |^2$.
The corresponding corrections, $\delta M^{ud}_V$ and $\delta Z^{ud}_V$, can be respectively extracted from the ``slope'' and the ``intercept'' of the ratio $\delta V^{ud} (t) / V^{ud} (t)$ at large time distances (see Refs.~\cite{deDivitiis:2011eh,deDivitiis:2013xla,Giusti:2017dmp}).
We have checked that the sum of the two terms in the r.h.s.~of Eq.~(\ref{eq:decomposition}) is independent of the specific choice of the value of $T_{data}$.

\section{Simulation details}
\label{sec:simulations}

The ETMC gauge ensembles used in this work are the same adopted in Ref.~\cite{Carrasco:2014cwa} to determine the up-, down-, strange- and charm-quark masses in isospin symmetric $QCD$ with $N_f = 2 + 1 + 1$ dynamical quarks.
The gauge fields are simulated using the Iwasaki gluon action~\cite{Iwasaki:1985we}. In the light-quark sector a unitary setup is adopted and the Wilson Twisted Mass action~\cite{Frezzotti:2000nk,Frezzotti:2003ni,Frezzotti:2003xj} at maximal twist is employed.
We consider three values of the inverse bare lattice coupling $\beta$, corresponding to lattice spacings varying from $0.089$ to $0.062 ~ \mbox{fm}$, pion masses in the range $M_\pi \simeq 220 \div 490 ~ \mbox{MeV}$ and different lattice volumes.
For earlier investigations of finite volume effects (FVEs) the ETMC had produced three dedicated ensembles, A40.20, A40.24 and A40.32, which share the same quark mass (corresponding to $M_\pi \simeq 320 ~ \mbox{MeV}$) and lattice spacing $(a \simeq 0.09 ~ \mbox{fm})$ and differ only in the lattice size $L$ $(L/a = 20 \div 32)$.
To improve such an investigation a further gauge ensemble, A40.40, has been generated at a larger value of the lattice size, $L/a = 40$.

The evaluation of the vector correlator has been carried out using the following lattice definition of the vector current:
\be
J_\mu^f(x) = Z_A ~ q_f ~ \overline{\psi}_{f^\prime} (x) \gamma_\mu \psi_f (x) ~ ,
\label{eq:localV}
\ee
where $\overline{\psi}_{f^\prime}$ and $\psi_f$ represent two quark fields with the same mass, charge and flavor, but regularized with opposite values of the Wilson $r$-parameter, {\it i.e.} $r_{f^\prime} = - r_f$.
Being at maximal twist, the current~(\ref{eq:localV}) renormalizes multiplicatively with the renormalization constant (RC) $Z_A$ of the axial current.
The local current~(\ref{eq:localV}) does not generate disconnected contributions in the vector correlator~(\ref{eq:VV}).
In addition, as discussed in Ref.~\cite{Giusti:2017jof}, the properties of the kernel function $f(t)$ in Eq.~(\ref{eq:delta_amu_t}), guarantee that the contact terms, generated in the HVP tensor by a local vector current, do not contribute to the evaluation of the muon anomalous magnetic moment.

The good statistical accuracy of the meson correlator relies on the use of the so-called ``one-end'' stochastic method~\cite{McNeile:2006bz}, which includes spatial stochastic sources at a single time slice chosen randomly.
We have calculated the vector correlator~(\ref{eq:VV}) adopting the local current~(\ref{eq:localV}) for the light $u$ and $d$ quarks using 160 stochastic sources (diagonal in the spin variable and dense in the color one) per each gauge configuration.

\section{Isospin-breaking corrections}
\label{sec:IB}

Let's now turn to the IB corrections $\delta V^{ud} (t)$ to the vector correlator at leading order in the small parameters $\alpha_{em}$ and $(\widehat{m}_d - \widehat{m}_u)$, which consist of the em, $\delta V^{QED} (t)$, and $SU(2)$-breaking, $\delta V^{QCD} (t)$, contributions.
The separation between the $QED$ and the strong IB terms is prescription dependent. In this contribution we impose a specific matching condition in which the renormalized couplings and quark masses in the full theory and in isosymmetric $QCD$ coincide in the $\overline{\rm MS}$ scheme at a scale of $2~\mbox{GeV}$~\cite{deDivitiis:2013xla,Gasser:2003hk}.

Using the expansion method of Refs.~\cite{deDivitiis:2011eh,deDivitiis:2013xla}, the em corrections $\delta V^{QED} (t)$ to the light-quark vector correlator can be computed through the evaluation of the self-energy, exchange, tadpole, pseudoscalar and scalar insertion diagrams described in Ref.~\cite{Giusti:2017jof}\footnote{The standard  combinatorial factor $1 / 2$ was missing in Eq.~(23) of Ref.~\cite{Giusti:2017dmp} and in Eq. (5.2) of Ref.~\cite{Giusti:2017jof}, but included in the calculations.}.
The removal of the photon zero-mode is done according to $QED_L$~\cite{Hayakawa:2008an}, i.e.~the photon field $A_\mu$ satisfies $A_\mu(k_0, \vec{k} = \vec{0}) \equiv 0$ for all $k_0$. We also adopt the quenched-$QED$ approximation, which treats the sea quarks as electrically neutral particles.
In addition one has to consider the $QED$ corrections to the RC of the vector current of Eq.~(\ref{eq:localV}), namely
\be
{\cal Z}_A = Z_A \left [ 1 + Z_A^{em} ~ Z_A^{fact} \right ] + {\cal O}(\alpha_{em}^2) ~ ,
\label{eq:ZAem}
\ee
where $Z_A$ is the RC of the current in pure $QCD$ (determined in Ref.~\cite{Carrasco:2014cwa}), $Z_A^{em}$ is the one-loop perturbative estimate of the $QED$ corrections at order ${\cal O}(\alpha_s^0)$ and $Z_A^{fact}$ takes into account corrections of order ${\cal O}(\alpha_{em} \alpha_s^n)$ with $n \geq 1$, i.e.~corrections to the ``naive factorization'' approximation in which $Z_A^{fact} = 1$.
We make use of the result $Z_A^{em} = - 15.7963 ~ \alpha_{em} ~ q_f^2 / (4 \pi)$ from Ref.~\cite{Martinelli:1982mw} and of a preliminary non-perturbative estimate~\cite{DiCarlo:PRD18} determined via the RI$^\prime$-MOM method, $Z_A^{fact} = 0.95 ~ (5)$, which improves the one obtained through the axial Ward-Takahashi identity derived in Ref.~\cite{Giusti:2017jof}.

\begin{SCfigure}[1.0]
\includegraphics[scale=0.4]{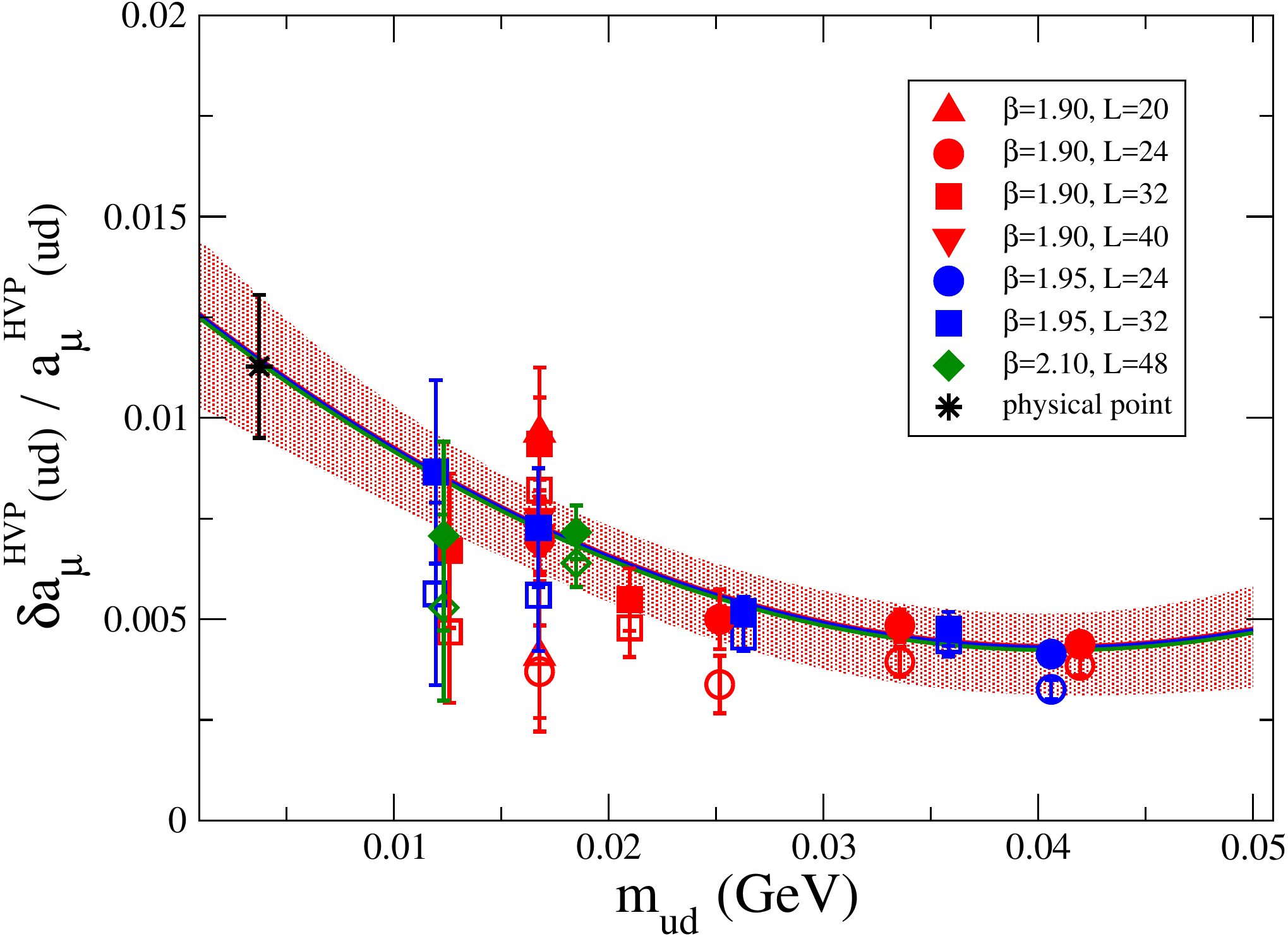}
\caption{\it \footnotesize Results for the ratio $\delta a_\mu^{\rm HVP} (ud) / a_\mu^{\rm HVP} (ud)$ versus the renormalized light-quark mass $m_{ud}$. The empty markers correspond to the raw data, while the full ones represent the lattice data corrected by the FVEs obtained in the fitting procedure~(\ref{eq:fit_ansatz}). The solid lines correspond to the results of the combined fit~(\ref{eq:fit_ansatz}) obtained in the infinite-volume limit at each value of the lattice spacing. The black asterisk represents the value of the ratio $\delta a_\mu^{\rm HVP} (ud) / a_\mu^{\rm HVP} (ud)$ extrapolated to the physical pion mass [corresponding to $m^{phys}_{ud} (\overline{\rm MS}, 2 ~ \rm GeV) = 3.70 ~ (17) ~ \rm MeV$, (determined in Ref.~\cite{Carrasco:2014cwa})] and to the continuum limit, while the red area indicates the corresponding uncertainty as a function of $m_{ud}$ at the level of one standard deviation. Plotted errors are statistical only.}
\label{fig:ratio_ud_IB_RM123}
\end{SCfigure}

The overall IB corrections $\delta a_\mu^{\rm HVP} (ud)$ can be calculated by summing the $QED$ and $QCD$ contributions in Eq.~(\ref{eq:delta_amu_t}).
The accuracy of the lattice data can be improved by forming the ratio of the IB corrections over the leading-order term.
Therefore, we have performed our analysis of the ratio $\delta a_\mu^{\rm HVP} (ud) / a_\mu^{\rm HVP} (ud)$, which is shown in Fig.~\ref{fig:ratio_ud_IB_RM123}.

\noindent We have performed combined extrapolations to the physical pion mass and to the continuum and infinite-volume limits adopting the following fitting function:
\be
\frac{\delta a_\mu^{\rm HVP} (ud)}{a_\mu^{\rm HVP} (ud)} = \delta_0 \left [ 1 + \delta_1 ~ m_{ud} + \delta_2 ~ {\rm X} + D ~ a^2 + FVE \right] ~ ,
\label{eq:fit_ansatz}
\ee
where we have considered both a quadratic $({\rm X} = m_{ud}^2)$ and a logarithmic $({\rm X} = m_{ud} ~ {\rm ln}(m_{ud}))$ phenomenological chiral dependence of the lattice data in order to estimate the systematic uncertainty due to the chiral extrapolation to the physical pion mass.
The finite-volume correction is estimated by using alternately one of the following fitting ansatzes\footnote{As previously observed in Ref.~\cite{Giusti:2017jof}, for neutral mesons with vanishing charge radius $QED$ FVEs are expected to start at order ${\cal O} (1 / L^3)$.}:
\bea
FVE & = & F ~ e^{- \bar M L} ~, \nonumber \\
FVE & = & \widehat{F}_n ~ \frac{\bar M^2}{16 \pi^2 f_0^2} ~ \frac{e^{- \bar M L}}{(\bar M L)^n}~, \quad\quad {\rm with} ~~ n = \frac{1}{2}, ~ 1, ~ \frac{3}{2}, ~ 2 \nonumber \\
FVE & = & \frac{\widetilde{F}}{L^3} ~,
\eea
where $\bar M^2 \equiv 2 B_0 m_{ud}$, $B_0$ and $f_0$ are the $QCD$ low-energy constants at leading order.
Discretization effects play a minor role and, for our ${\cal O} (a)$-improved simulation setup, they can be estimated by including $(D \neq 0)$ or excluding $(D = 0)$ the term proportional to $a^2$ in Eq.~(\ref{eq:fit_ansatz}).
The free parameters to be determined by the fitting procedure are $\delta_0, ~ \delta_1, ~ \delta_2, ~ D$ and $F (\widehat{F}_n, ~ \widetilde{F})$.

In our combined fit the values of the parameters are determined by a $\chi^2$-minimization procedure adopting an uncorrelated $\chi^2$.
The uncertainties on the fitting parameters do not depend on the $\chi^2$-value, because they are obtained by using the bootstrap samplings of Ref.~\cite{Carrasco:2014cwa}.
This guarantees that all the correlations among the lattice data points and among the fitting parameters are properly taken into account.
As for the lattice spacing $a$ and the RCs $Z_P$, which enters in the definition of the renormalized quark mass, their uncertainties are taken into account by imposing a Gaussian prior in the fitting procedure.

At the physical pion mass and in the continuum and infinite-volume limits we get the preliminary result
\be
\frac{\delta a_\mu^{\rm HVP} (ud)}{a_\mu^{\rm HVP} (ud)} = 0.011 ~ (3)_{stat+fit+input} (2)_{chir} (2)_{FVE} (1)_{disc} (1)_{{\cal Z}_A} ~ [4] ~ ,
\label{eq:ratio_IB}
\ee
where the errors come in the order from (statistics + fitting procedure + input parameters of the eight branches of the quark mass analysis of Ref.~\cite{Carrasco:2014cwa}), chiral extrapolation, finite-volume and discretization effects and from the uncertainty on the $QED$ contribution to the RC $Z_A$ (see Eq.~(\ref{eq:ZAem})).
In Eq.~(\ref{eq:ratio_IB}) the uncertainty in square brackets corresponds to the sum in quadrature of the statistical and systematic errors.
Further details will be given in a forthcoming paper, including a dedicated study of FVEs as well as an estimate of the systematic uncertainty related to the quenched-$QED$ approximation.

Using the leading-order result of Ref.~\cite{Giusti:2018mdh}, $a^{\rm HVP}_\mu (ud) = 619.0 ~ (17.8) \cdot 10^{-10}$, we obtain a preliminary determination of the leading-order IB corrections to $a^{\rm HVP}_\mu (ud)$, namely
\be
\delta a^{\rm HVP}_\mu (ud) = 7 ~ (2) \cdot 10^{-10} ~ ,
\label{eq:result_ud_IB}
\ee
which is the sum of the em contribution,
\be
\left [ \delta a^{\rm HVP}_\mu (QED) \right ] (\overline{\rm MS}, 2 ~ \mbox{GeV}) = 1.3 ~ (0.9) \cdot 10^{-10} ~,
\ee
and of the strong IB one,
\be
\left [ \delta a^{\rm HVP}_\mu (QCD) \right ] (\overline{\rm MS}, 2 ~ \mbox{GeV}) = 5.6 ~ (2.0) \cdot 10^{-10} ~.
\ee
We emphasize that the IB corrections~(\ref{eq:result_ud_IB}) are by far dominated by the $SU(2)$-breaking term, which is roughly of the order of $\sim 80\%$ of $\delta a^{\rm HVP}_\mu (ud)$.
In addition, we point out that the bulk of the IB corrections to $a^{\rm HVP}_\mu$, $a^{\rm HVP}_\mu (IB)$, comes from the light-quark contribution, being the strange and charm-quark ones negligible with respect to the uncertainties of the leading-order terms~\cite{Giusti:2017jof}.
Our lattice determination~(\ref{eq:result_ud_IB}), obtained with $N_f = 2 + 1 + 1$ dynamical flavors of sea quarks, agrees within the errors with and is more precise than the phenomenological estimate of Ref.~\cite{Borsanyi:2017zdw}, $\delta a^{\rm HVP}_\mu (ud) = 7.8 ~ (5.1) \cdot 10^{-10}$, by the BMW Collaboration and the lattice determination of the RBC/UKQCD Collaborations, $\delta a^{\rm HVP}_\mu (ud) = 9.5 ~ (10.2) \cdot 10^{-10}$,~\cite{Blum:2018mom} $(N_f = 2 + 1)$.
Furthermore, the FNAL/HPQCD/MILC Collaborations have recently determined the strong IB contribution, $\delta a^{\rm HVP}_\mu (QCD) = 9.0 ~ (4.5) \cdot 10^{-10}$,~\cite{Chakraborty:2017tqp} $(N_f = 1 + 1 + 1 + 1)$.

Adding the contributions of light, strange and charm quarks, $a^{\rm HVP}_\mu (ud) = 619.0 ~ (17.8) \cdot 10^{-10}$, $a^{\rm HVP}_\mu (s) = 53.1 ~ (2.5) \cdot 10^{-10}$ and $a^{\rm HVP}_\mu (c) = 14.75 ~ (0.56) \cdot 10^{-10}$, determined by the ETMC in Refs.~\cite{Giusti:2017jof} and~\cite{Giusti:2018mdh}, our preliminary result for the IB corrections, $a^{\rm HVP}_\mu (IB) = 7 ~ (2) \cdot 10^{-10}$, and an estimate of the quark-disconnected diagrams, $a^{\rm HVP}_\mu (disconn.) = -12 ~ (4) \cdot 10^{-10}$, obtained using the results of Refs.~\cite{Borsanyi:2017zdw} and~\cite{Blum:2018mom}, we get
\be
a^{\rm HVP}_\mu = 682 ~ (19) \cdot 10^{-10} ~ ,
\ee
which agrees with the recent determinations based on dispersive analyses of the experimental cross section data for $e^+ e^-$ annihilation into hadrons (see {\it e.g.} Ref.~\cite{Keshavarzi:2018mgv} and references therein).

\section*{Acknowledgments}
We warmly thank our colleagues R.~Frezzotti and N.~Tantalo for enjoyable discussions. We gratefully acknowledge the CPU time provided by CINECA under the initiative INFN-LQCD123 on the KNL system Marconi at CINECA (Italy).

\end{document}